\documentclass[reprint,]{revtex4-2}
\usepackage{color,amsmath,amssymb,commath,graphicx}

\begin{document}

\title{Tradeoffs between energy efficiency and mechanical response in fluid flow networks}
\author{Sean Fancher}
\email{sfancher@sas.upenn.edu}
\affiliation{Department of Physics and Astronomy, University of Pennsylvania, Philadelphia, PA 19104, USA}
\author{Eleni Katifori}
\affiliation{Department of Physics and Astronomy, University of Pennsylvania, Philadelphia, PA 19104, USA}

%Editors: Herbert Levine, Curtis Callan, Mehran Kardar
%NAS members: William Bialek, Marsha Berger, Boris Shraiman, Nigel Goldenfeld, Eugene Stanley
%Referees: Dirk Whitthaut, Doug Kelley, Jordi Alastruey

\begin{abstract}
Transport networks are typically optimized, either by evolutionary pressures in biological systems or by human design in engineered structures. In the case of systems such as the animal vasculature, the transport of fluids is hindered by the inherent viscous resistance to flow while being kept in a dynamic state by the pulsatile nature of the heart and elastic properties of the vessel walls. While this imparted pulsatility necessarily increases the dissipation of energy caused by the resistance, the vessel elasticity helps to reduce overall dissipation by attenuating the amplitude of the pulsatile components of the flow. However, we find that this reduction in energy loss comes at the price of increasing the time required to respond to changes in the flow boundary conditions for vessels longer than a critical size. In this regime, dissipation and response time are found to follow a simple power law scaling relation both in single vessels as well as hierarchically structured networks. We validate the model in human vasculature and apply biologically relevant parameters to show that response time has to be considered alongside dissipation as an important fitness cost function in the evolutionary optimization of animal vascular networks.
\end{abstract}

\maketitle

\section{Introduction}
\label{sec:intro}

The material transport network is a ubiquitous model that can be used to describe a large array of naturally and artificially occurring systems. The various properties of such networks are frequently either thought to be adapted through natural selection \cite{lucas2013plant,monahan2013evolutionary} or specifically designed \cite{gallego2001optimal,jimenez2010determination,ziari2012optimal} to optimize certain aspects of the local flow of material or the global performance of the entire network. One particularly common optimized quantity is the rate at which the energy to maintain the flow is dissipated throughout the network. Minimizing dissipation for a certain delivered flow can be constrained, however, by other costs to construct or adapt the network such as the metabolic cost to maintain the vessels, the requirement for resilience to damage, and more \cite{murray1926physiological,sherman1981connecting,Katifori2010a,Postnov2016,Ronellenfitsch2016,Tekin2016,Ronellenfitsch2019,Kirkegaard2020,Chang2019,Planck2019,Erlich2019}.

One easily identifiable factor that complicates minimizing dissipation is pulsatility in the driving force that maintains the flow of material. For systems such as fluid undergoing laminar flow or electrical current traversing a linear resistor, the rate of energy dissipation per unit length along a vessel is proportional to the square of the local current. In a pulsatile system with a well defined period, the time-averaged value of the square of the current will necessarily be greater than the square of the average, meaning the inclusion of any pulsatile components to the flow can only cause the time-averaged dissipation to increase. Thus, such linear systems utilizing a pulsatile driving force will necessarily have greater total dissipation over the course of a full period than an equivalent system utilizing a continuous, steady flow pump with the same average flow rate.

\begin{figure}[t]
    \centering
    \includegraphics[width=\columnwidth]{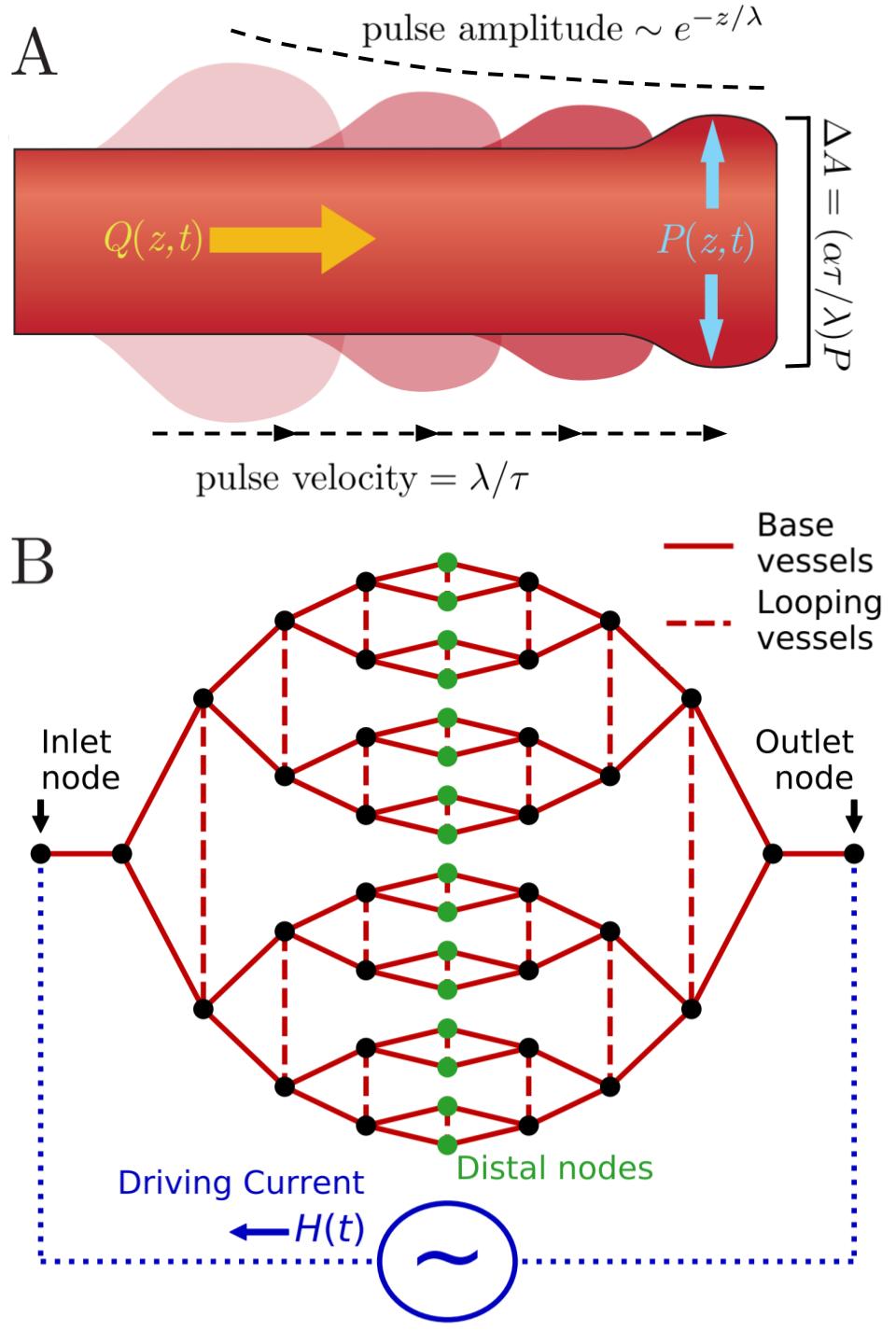}
    \caption{A) Fluid flowing with volumetric rate $Q$ through a compliant vessel causes the vessel to expand and contract in response to changes in pressure, $P$. Vessel parameters $\lambda$, $\tau$, and $\alpha$ dictate the pulse decay length scale, pulse velocity, and deformation size of the cross sectional area, $A$. B) Such vessels can be combined into a linear flow network in which each individual vessel has its own distinct parameter values. Here, we consider as a toy model a 4 generation bifurcating and recombining tree powered by an external flow condition, $H$, at the inlet and outlet nodes. We also consider a similar network in which looping vessels form cross connections between generations.}
    \label{diagrams}
\end{figure}

Despite this drawback, many transport networks rely on pulsatile drivers to maintain the flow of material. One common such example is the animal vascular network. In humans, blood pressure and flow have pulsatile components imparted by the periodic beating of the heart. The energy of the pulse is stored within the kinetic momentum of the blood itself and the elastic energy due to the nonzero compliance of arterial walls, resulting in traveling waves in flow. However, the amplitude of this pulsatility has been observed to decrease as the blood is transported further away from the heart \cite{jones1971effects,hashimoto2014central}. Damping of pulsatility can be attributed to the inherent resistance in the flow of blood dissipating this energy, resulting in pressure that attenuates as depicted in Fig. \ref{diagrams}A \cite{holenstein1988reverse,sherwin2003one,alastruey2012physical,flores2016novel,yigit2016non}. The culmination of these effects eventually allows the heart to maintain the same mean flow with less overall energy expenditure than it would without damping of pulsation, which is evolutionarily advantageous.

Pulsatility is not absolutely damped, however, as it remains to some degree even at the capillary level \cite{harazny2014first}. We might then wonder why evolution did not do away with pulsatility entirely, by developing ways to remove it altogether. One such obvious reason is that removing pulsatility entirely will require a vessel architecture that is fundamentally incompatible with what can be built biologically. In this work, we posit the existence of another effect, an increase in the network's response time, which might prevent evolution from further decreasing pulsatility and the energy necessary to maintain a certain mean blood flow.

Vessel compliance in particular has been shown to affect the extent of pulsatile damping, with stiffer, less compliant vessels leading to weaker damping \cite{holenstein1988reverse,bui2009dynamics}. While several theoretical and modelling studies have investigated the effects of vessel compliance on pressure waveforms and energetic costs in subsections of biological fluid flow networks such as the major arteries \cite{holenstein1988reverse,sherwin2003one,bui2009dynamics,alastruey2012physical,pan2014one,flores2016novel,perdikaris2015effective,yigit2016non,bauerle2020living}, exactly how the effects of compliance might alter the global properties of the entire vascular network, particularly when the pulsatile driver relinquishes its periodicity to transition between different steady states, remains poorly understood. Such an understanding could provide deeper insight into the evolutionary pressures that formed systems like the animal vasculature as well as more reliable models for generating artificial networks or prosthetic devices.

Here, we investigate the effects of vessel compliance on the material flow in transport networks with time-dependent boundary conditions. In Sec. \ref{sec:scalings} we utilize a linear flow model developed previously \cite{fancher2022mechanical} to derive the rate of energy dissipation within a transport vessel under periodic flow conditions. We find that while increasing the compliance of the vessels within a network does decrease the rate of energy dissipation whenever pulsatility is present, it also has the consequence of increasing the amount of time required for the system to adapt to changes in the boundary conditions, such as a change in the heart rate of a vascular system. These effects follow a simple power law relation for vessel sizes above a critical, compliance dependent value. We show that this result holds for single vessels, for which the equations can be solved analytically, as well as whole networks of the form depicted in Fig. \ref{diagrams}B, for which we use numerical integration. In Sec. \ref{sec:app} we apply our methodology to a model of the entire human vasculature adapted from \cite{mynard2015one} and show that global multiplicative changes to vessel compliance induce the exact same scaling behavior between dissipation and response time. Our work highlights the importance of this response time, the time for the network to adapt to the new pump flow conditions, as an important design consideration for networks composed of elastic vessels.

\section{Results}
\label{sec:results}

\subsection{Dissipation and response scaling functions}
\label{sec:scalings}

We begin by analyzing the mechanics of fluid flowing through a single vessel comprised of compliant walls via a model which has been developed in a previous work \cite{fancher2022mechanical}. Specifically, we consider the volumetric flow rate, $Q(z,t)$, and fluid pressure, $P(z,t)$, where $z$ is the spatial dimension along the vessel axis, $t$ is time, and $P$ has been averaged over the cross sectional area at position $z$. To obtain the dynamic equations for $Q$ and $P$ we enforce that the fluid be incompressible and linearize the Navier-Stokes equation under a set of assumptions; namely that the system is rotationally symmetric, the Womersley number is sufficiently small that the fluid velocity profile is approximately that of Poiseuille flow, the flow velocity is small compared to the velocity of current and/or pressure pulses, the wavelength and/or exponential length scale of the flow profile is large compared to vessel radius, and changes to the vessel cross sectional area are small and approximately linear with changes in pressure. Under these enforcements and assumptions it can be shown that $Q$ and $P$ must obey

\begin{equation}
    \lambda\frac{\partial Q}{\partial z}+\alpha\tau\frac{\partial P}{\partial t} = 0,
    \label{Qdyn}
\end{equation}

\begin{equation}
    \alpha\lambda\frac{\partial P}{\partial z}+\tau\frac{\partial Q}{\partial t}+2Q = 0.
    \label{Pdyn}
\end{equation}

Eqs. \ref{Qdyn} and \ref{Pdyn} are expressed in terms of the characteristic length scale $\lambda$, time scale $\tau$, and admittance $\alpha$. As shown in Fig. \ref{diagrams}A, these dictate the decay length and velocity of current and/or pressure pulses as they travel through the vessel as well as the cross sectional area deformation magnitude. These characteristic parameters can also be reexpressed in terms of the physical properties given by the flow resistance per unit length $r$, fluid interia $\ell$, and vessel wall compliance per unit length $c$ via $\lambda=2\sqrt{\ell/c}/r$, $\tau=2\ell/r$, and $\alpha=\sqrt{c/\ell}$.

With the dynamic equations known, we now investigate the rate of energy dissipation over a single vessel of total length $L$. Assuming the flow boundary conditions, $Q(0,t)$ and $Q(L,t)$, are known, Eqs. \ref{Qdyn} and \ref{Pdyn} can be solved exactly up to an additive constant in $P$ which must be determined by choice of gauge. Here, we consider the case in which the boundary conditions are symmetric ($Q(0,t)=Q(L,t)$) and pulsatile with a well defined period $T$ and corresponding angular frequency $\omega=2\pi/T$. Pulsatility allows the flow to be expanded into a Fourier series of the form $Q(z,t)=\sum_{n}\tilde{Q}^{(n)}(z)\text{exp}(in\omega t)$, where the summation is over all integers. The pressure can also be expanded in this way with similar notation. The imposed symmetry allows the total time averaged dissipation rate, $\langle D\rangle=T^{-1}\int_{0}^{T}dt\int_{0}^{L}dz\>rQ^{2}\left(z,t\right)$, to be expressed as $rL\sum_{n}|\tilde{Q}^{(n)}(0)|^{2}f(L/\lambda,n\omega\tau)$ where

\begin{equation}
    f(x,y) = \frac{k_{\text{I}}\left(y\right)\sinh\left(xk_{\text{R}}\left(y\right)\right)+k_{\text{R}}\left(y\right)\sin\left(xk_{\text{I}}\left(y\right)\right)}{xy\left(\cosh\left(xk_{\text{R}}\left(y\right)\right)+\cos\left(xk_{\text{I}}\left(y\right)\right)\right)},
    \label{DSF}
\end{equation}

\noindent and $k_{\text{R}}(y)$ and $k_{\text{I}}(y)$ are the real and imaginary parts of $k(y)=\sqrt{iy(2+iy)}$ respectively, with the principle root being taken to ensure $k_{R}(y)\ge 0$ for $y\in\mathbb{R}$. We denote the function $f(L/\lambda,n\omega\tau)$ as the \textit{dissipation scaling function} for the chosen system under the given constraints and derive it in the Supplement. Specifically, the dissipation scaling function dictates the factor by which individual Fourier components of the flow deviate from the baseline dissipation of $rL|\tilde{Q}^{(n)}|^{2}$.

\begin{figure}[t]
    \centering
    \includegraphics[width=\columnwidth]{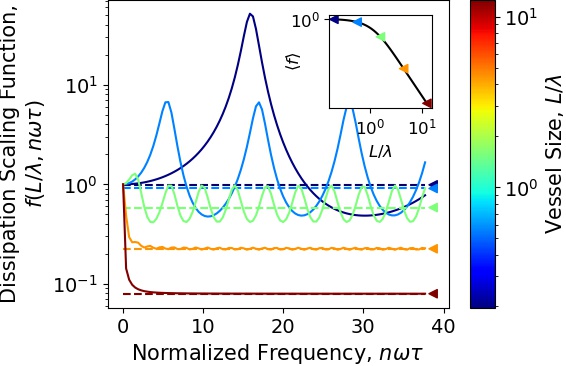}
    \caption{Plot of dissipation scaling function (Eq. \ref{DSF}) as a function of frequency for a variety of vessel lengths. The mean dissipation is given by the dashed lines in the main plot and black curve in the inset. Triangle markers in both plots also designate the value of the mean dissipation for each vessel length shown. While short vessels experience large fluctuations in dissipation values, the frequency averaged dissipation (Eq. \ref{DSFmean}) monotonically decreases with increasing vessel size.}
    \label{SVDSF}
\end{figure}

We can examine the effects the various parameters have on the overall dissipation rate by observing the properties of the dissipation scaling function, as shown in Fig. \ref{SVDSF}. Specifically, we are most interested in the effects of varying the compliance, $c$, while holding fixed the resistance, $r$, and interia, $\ell$, to maintain a constant baseline dissipation and fluid density respectively. In terms of the characteristic parameters $\lambda$, $\tau$, and $\alpha$, this is equivalent to holding $\tau$ and the product $\alpha\lambda$ constant. Since $\alpha$ is absent from the expression $f(L/\lambda,n\omega\tau)$ though, we can simply focus on the effects of varying $\lambda$.

From Fig. \ref{SVDSF} we see that the dissipation scaling function fluctuates periodically in frequency space but tends to decrease as $L/\lambda$ is increased, excepting regions of overlapping resonant and antiresonant frequencies. These fluctuations can be integrated out by defining the frequency averaged dissipation scaling function via the operation

\begin{equation}
    \left\langle f\left(\frac{L}{\lambda}\right)\right\rangle^{-1} = \lim_{\Omega\to\infty}\frac{1}{\Omega}\int_{0}^{\Omega}d\omega\>\left(f\left(\frac{L}{\lambda},\omega\tau\right)\right)^{-1},
    \label{DSFmean}
\end{equation}

\noindent which explicitly evaluates to $\langle f(L/\lambda)\rangle=\tanh(L/\lambda)/(L/\lambda)$ under Eq. \ref{DSF} (Fig. \ref{SVDSF} inset). Under the assumption that the dissipation scaling function can be sufficiently approximated by this value, the total dissipation can be taken to scale as $rL\langle f(L/\lambda)\rangle$. In the case of symmetric flow boundary conditions considered here, this implies that the total dissipation scales linearly with the total resistance for short vessels since $rL\langle f(L/\lambda)\rangle\approx rL$ when $L\ll\lambda$. This reflects the simple fact that the current does not appreciably change over the span of a short vessel, allowing the dissipation to be well approximated by $rL|\tilde{Q}^{(n)}(0)|^{2}$. Conversely, for long vessels the total dissipation becomes independent of $L$ and scales linearly with $\lambda$ since $rL\langle f(L/\lambda)\rangle\approx r\lambda$ when $L\gg\lambda$. This is due to the fact that $\lambda$ dictates the distance over which the flow and pressure waves are damped before becoming negligible and thus cease to contribute meaningfully to the dissipation. These scaling results are valid in the limit of high frequencies, but for $\omega\rightarrow 0$, the dissipation scaling function limits to a value of 1 and the total dissipation simplifies to that of typical Poiseuille flow viscous losses.

With an understanding of how compliance affects dissipation via its relation to the characteristic parameters $\lambda$, $\tau$, and $\alpha$, we now focus on the dynamic properties of the flow. In particular, it has been noted in computational models that an increase in vessel compliance correspondingly causes an increase in the time required to establish a new steady state when the boundary conditions change \cite{bui2009dynamics}. We have previously investigated the precise scaling of this response time in terms of the system parameters \cite{fancher2022mechanical} and found that when the boundary conditions undergo a sudden shift of the form $Q(0,t)=Q(L,t)=H(t)+\Delta H(t)\Theta(t)$ for arbitrary pulsatile functions $H(t)$ and $\Delta H(t)$, the flow and pressure exponentially approach their post-transition steady state as $\text{exp}(-t/(\tau\beta(L/\lambda)))$, where

\begin{equation}
    \beta\left(\frac{L}{\lambda}\right) = \begin{cases}
    1 & \frac{L}{\lambda}\le\pi \\
    \left(1-\sqrt{1-\left(\frac{\pi\lambda}{L}\right)^{2}}\right)^{-1} & \frac{L}{\lambda}>\pi \end{cases}.
    \label{betadef}
\end{equation}

\begin{figure*}[t]
    \centering
    \includegraphics[width=\textwidth]{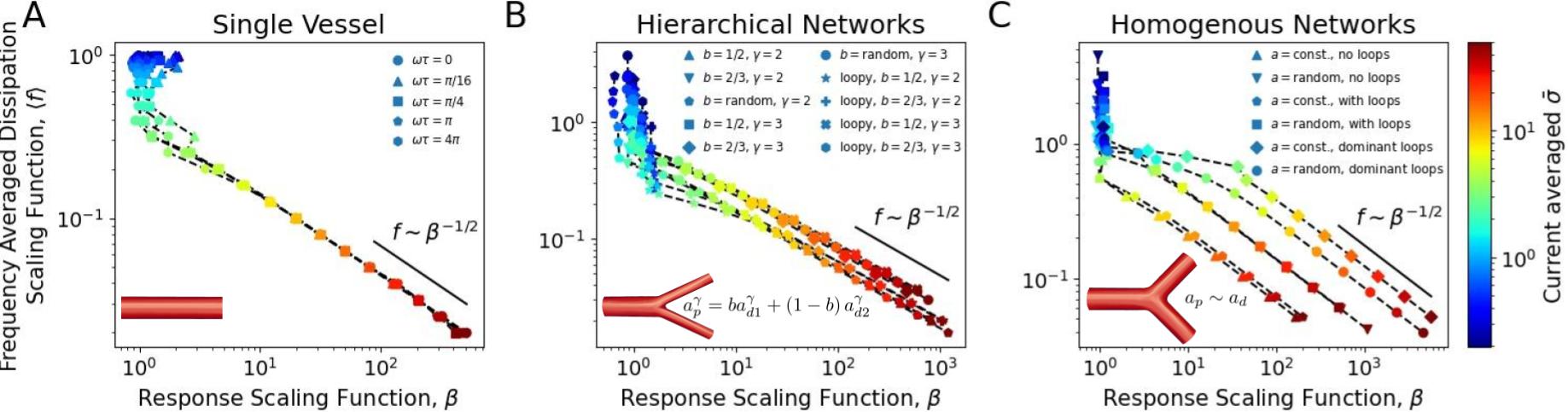}
    \caption{Plots of the mean dissipation scaling function, $\langle f\rangle$, against the response scaling function, $\beta$. A) In the single vessel case where $\bar{\sigma}=L/\lambda$, $\langle f\rangle$ and $\beta$ are given explicitly by Eqs. \ref{DSFmean} and \ref{betadef}. For short vessels ($L/\lambda<\pi$), the dissipation is seen to slowly decrease with vessel size while the response time remains constant. For large vessels ($L/\lambda>\pi$), the dissipation scales as the inverse square root of the response time. B) Hierarchical networks show the same qualitative behaviors when dissipation and response time are expressed in terms of the path averaged decay factor, $\bar{\sigma}$. C) Homogeneous networks also exhibit the same qualitative behaviors, but with more drawn out transition regions when vessel sizes are randomized or large looping vessels are introduced.}
    \label{DVR}
\end{figure*}

The function $\beta(L/\lambda)$ is identified as the response scaling function and dictates the time scale of the approach to steady state in units of $\tau$ (see \cite{fancher2022mechanical} for explicit derivation and numeric verification). Importantly, $\beta(L/\lambda)$ is independent of frequency but dependent on the vessel size, $L/\lambda$. This allows for a direct comparison to the frequency averaged dissipation scaling function given by Eq. \ref{DSFmean}, which Fig. \ref{DVR}A plots against the numerically derived response time values for a variety of different vessel sizes in both the nonpulsatile and pulsatile boundary current cases. In the short vessel regime ($L/\lambda<\pi$), the dissipation scaling function decreases weakly with vessel size while the response scaling function is simply constant. In the long vessel regime ($L/\lambda >\pi$), the dissipation and response scaling functions are well approximated as power laws of the form $\langle f(L/\lambda)\rangle\approx (L/\lambda)^{-1}$ and $\beta (L/\lambda)\approx (2/\pi^{2})(L/\lambda)^{2}$ respectively, thus producing the relative scaling relation $f\sim\beta^{-1/2}$ shown in Fig. \ref{DVR}A. Since larger values of vessel compliance create larger values of $L/\lambda$, this scaling relation introduces a tradeoff wherein high vessel compliance allows for decreased dissipation due to pulsatile components being damped away but at the cost of the system being very slow to respond to changes in boundary conditions. 

We now look to examine this tradeoff in the context of entire networks of compliant vessels. For the time being we consider the simple toy network depicted in Fig. \ref{diagrams}B. This network is comprised of a pair of inlet and outlet vessels that systematically bifurcate into a series of base vessels until connecting at a collection of distal nodes. Each vessel is taken to have the same length, but the values of $\lambda$, $\tau$, and $\alpha$ are allowed to be chosen independently. We presently consider the case in which 4 generations of base vessels exist. A ``loopy" version of the network is then separately constructed by connecting intergenerational nodes via looping vessels. Finally, a symmetric driving current is imposed such that the currents flowing into the inlet node and out of the outlet node are equivalent. Mass conservation and pressure continuity are then enforced at each non-boundary node. These boundary conditions and connectivity laws mean that once the explicit form of the driving current and a pressure gauge are defined, the Fourier components of the current and pressure everywhere within the network can be solved for (see Supplement).

Using this method, we construct a variety of different networks. These are separated into two distinct groups: hierarchical and homogeneous. In the former case, the radius of each vessel, denoted as $a$, is dictated by the branching equation

\begin{equation}
    a_{\text{parent}}^{\gamma} = ba_{\text{daughter 1}}^{\gamma}+\left(1-b\right)a_{\text{daughter 2}}^{\gamma},
    \label{branch}
\end{equation}

\noindent where $b\in (0,1)$ is the branching ratio and $\gamma$ is the branching exponent. The radius of any looping vessel is taken to satisfy that the cross sectional area of that vessel is the arithmetic average of the cross sectional areas of the parent vessels leading into the nodes connected by the looping vessel. In all hierarchical networks considered here, $\gamma$ is held constant over the whole network while $b$ is either similarly held constant or chosen randomly from a triangular distribution between 0.05 and 0.95 at each bifurcation node. Thus, once the radius of the inlet and outlet vessels are specified, the radii of all other vessels are determined in this way.

For homogeneous networks, the radii of each vessel are either taken to be constant across the entire network or independently drawn from a log-normal distribution. The radii of looping vessels are determined in the same manner, though we also consider the case in which the looping vessels are made ``dominant" by increasing their radii by a factor of 10. This ``dominant loops" construction allows us to investigate the effects of vessels that do not significantly contribute to a nonpulsatile steady state but still carry a significant proportion of the fluid volume. In both the hierarchical and homogeneous constructions, the resulting cross sectional areas of each vessel are used to determine the values of $\lambda$, $\tau$, and $\alpha$ assigned to each vessel via the linear scaling of each parameter with said cross section \cite{fancher2022mechanical}.

Once the networks are fully constructed, the space and time scales that dictate the flow and pressure dynamics must be codified. Specifically, we note that in the single vessel case, only the values of $L/\lambda$ and $\tau$ were relevant in determining both the dissipation and response time of the system. To obtain a network equivalent to $L/\lambda$, we consider a single path through the network from the inlet to outlet node, denoted as $\mathcal{S}$, such that the flow in the traversal direction is always nonnegative when a constant, nonpulsatile driving current is imposed. We express the nondimensionalized length of this path as $\sigma_{\mathcal{S}}=\int_{\mathcal{S}}dz\>(1/\lambda(z))$, where $z$ denotes the positional variable along the path and $\lambda(z)$ is simply a function of which vessel the path resides in at position $z$. This definition simplifies to $L/\lambda$ in the single vessel case when $\lambda(z)$ is constant and the bounds of integration are simply $\lbrack 0,L\rbrack$. This nondimensional path length is then averaged over the entire network by summing over all such possible paths, $\bar{\sigma}=\sum_{\mathcal{S}}\sigma_{\mathcal{S}}Q_{\mathcal{S}}$, where $Q_{\mathcal{S}}$ is the normalized path weight determined by the proportion of fluid that traverses that path under the aforementioned nonpulsatile driving current. We can similarly obtain a network time scale by considering the parameter $\nu_{\mathcal{S}}=\sigma_{\mathcal{S}}^{-1}\int_{\mathcal{S}}dz\>(\tau(z)/\lambda(z))$, which represents the time required for a wavefront to traverse the chosen path normalized by the nondimensional path length. In the single vessel case, this is equivalent to the identity $\tau=(\tau L/\lambda)/(L/\lambda)$, while in the network case we can utilize the same averaging method to obtain $\bar{\nu}=\sum_{\mathcal{S}}\nu_{\mathcal{S}}Q_{\mathcal{S}}$. Thus, $\bar{\sigma}$ and $\bar{\nu}$ provide network averaged, spatially nondimensionalized length and time scales equivalent to $L/\lambda$ and $\tau$ in single vessels \cite{fancher2022mechanical}.

We now move on to calculating the dissipation and response scaling functions for each considered network. The network dissipation can be numerically calculated for any given driving current by inverting the network Laplacian matrix and summing over the Fourier modes (see Supplement). With this method, we can define the dissipation scaling function over frequency space as $f(\omega) = D(\omega)/D(0)$, where $D(0)$ is the dissipation in the network when a constant, nonpulsatile driving current of unit magnitude is imposed and $D(\omega\ne 0)$ is the dissipation when the driving current is of the form $H(t)=\sqrt{2}\cos(\omega t)$. This definition of the dissipation scaling function is notably dependent on the structure of the network and the values of $\lambda$, $\tau$, and $\alpha$ for each vessel therein while still reproducing the exact same form of $f(L/\lambda,n\omega\tau)$ obtained in the single vessel case. The frequency averaged value can also be obtained by numerically integrating the dissipation scaling function over frequency space in the same manner as given in Eq. \ref{DSFmean}.

The value of the response scaling function for each network was obtained in our previous work \cite{fancher2022mechanical} by simulating the evolution of current and pressure throughout the network, monitoring the cumulative flow through the distal nodes, and performing a linear fit to the residual current on a semilog scale with the time axis scaled such that $\bar{\nu}=1$. The response scaling function was taken to be equivalent to the negative inverse of the fitted slope. This process, along with the calculation of the dissipation scaling function described above, is repeated for a wide range of different $\bar{\sigma}$ values obtained by scaling the compliance of each vessel by a common factor, $c_{f}$, which causes $\bar{\sigma}\sim\sqrt{c_{f}}$ without affecting $\bar{\nu}$.

Fig. \ref{DVR}B and C plot the dissipation and response scaling functions against each other as functions of $\bar{\sigma}$ for hierarchical and homogeneous networks respectively. These plots clearly show that every network considered here displays the two primary qualitative features seen in the single vessel case: decreasing dissipation and constant response time at small $\bar{\sigma}$ and an inverse square root power law relation at large $\bar{\sigma}$. Intriguingly, the networks appear to form distinct clusters at large $\bar{\sigma}$ as well. Hierarchical networks are seen to separate based on the branching exponent, $\gamma$, with $\gamma=2$ networks displaying lower dissipation and higher response time than $\gamma=3$ networks for equivalent $\bar{\sigma}$ values. Meanwhile, homogeneous networks form three distinct groups wherein all networks have similar dissipation but response time undergoes significant increases between networks with constant vessel sizes, networks with random vessel size, and networks with dominant loops respectively.

\subsection{Application to the human vasculature}
\label{sec:app}

We now seek to apply the methodology presented here to the full human vasculature for the purpose of better understanding how vessel compliance affects dissipation and response time. To do so, we utilize the network constructed by Mynard and Smolich \cite{mynard2015one}. The Mynard-Smolich model itself uses a collection of nonlinear dynamical equations to simulate the flow and pressure through hundreds of vessels representing arteries, veins, and capillary beds in the systemic and pulmonary networks. The resulting {\it in silico} waveforms require significant computational power to generate but are seen to closely match experimental measurements.

Here, we use Eqs. \ref{Qdyn} and \ref{Pdyn} as the dynamical equations for flow and pressure to construct a linearized analogue of the Mynard-Smolich network. This is done by using data provided in the supplementary material of \cite{mynard2015one} to define the connectivity between the 127 arteries, 161 veins, and 27 capillary beds on the systemic side of the heart. We ignore the pulmonary network here, as it is functionally separated from the systemic network by the heart. The length of each artery and vein is also given directly by the supplementary data while the physical parameters ($\lambda$, $\tau$, and $\alpha$) are determined from the provided mean cross sectional area and properties of the blood and vessel material. The capillary beds are treated notably different from the arteries and veins. As was done in the Mynard-Smolich model, these are not considered to be full vessels but lumped parameter approximations of a collection of vessels. Specifically, each capillary bed is treated as a simple linear resistor connected in parallel to two separate grounded capacitors, one on each side. The resistance and capacitance of each of the components is again given directly by the supplementary data of the Mynard-Smolich model. From this construction, we can derive a network Laplacian in much the same way as was done for the networks considered in the previous section and perform a variety of {\it in silico} experiments based on our far less computationally demanding linear model (see Supplement).

\begin{figure*}[t]
    \centering
    \includegraphics[width=0.88\textwidth]{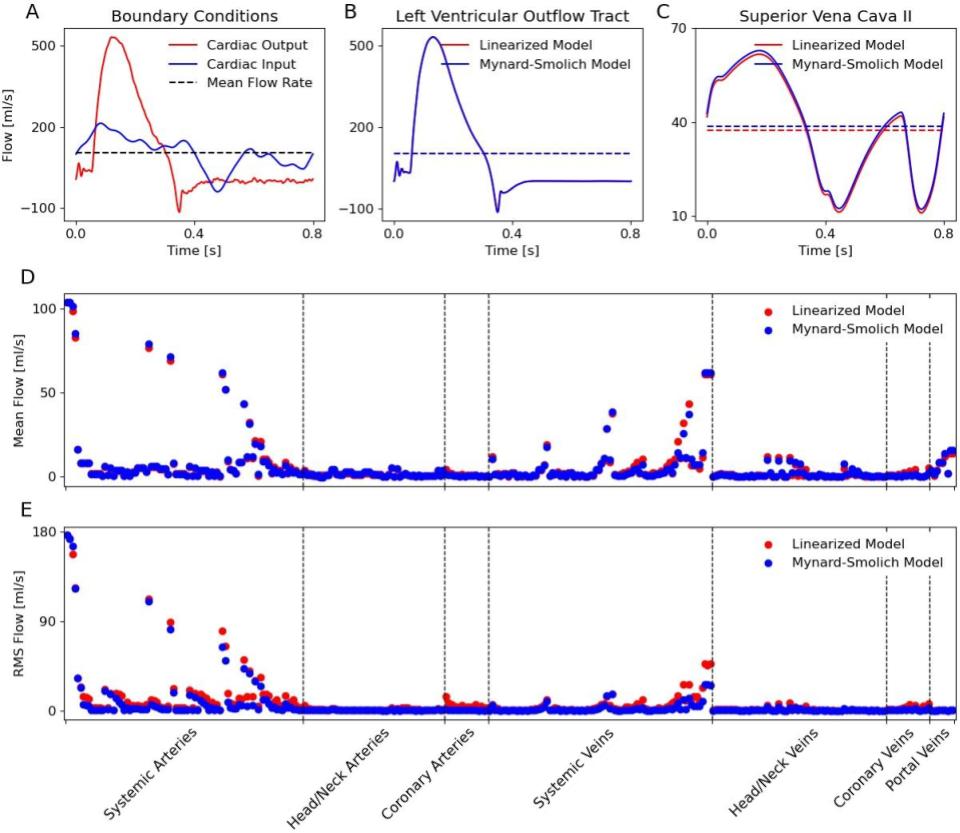}
    \caption{Comparing the linearized model investigated here with the nonlinear model developed by Mynard and Smolich \cite{mynard2015one}. After constructing the network, cardiac input and output waveforms (A) are derived so as to maximize agreement between the linearized model and data from the Mynard-Smolich model in the Left Ventricular Outflow Tract (B) and Superior Vena Cava II (C). These boundary conditions allow us to solve for the flow and pressure everywhere in the linearized model and compare the mean flow (D) and root-mean-square flow (E) within each individual vessel between the two models. The vessels are categorized based on their anatomical location and loosely sorted based on their distance from the heart along the $x$-axis of each plot.}
    \label{verfig}
\end{figure*}

First, we directly compare our linearized model to the Mynard-Smolich model. To do so, we establish two boundary conditions; the flow forced out of the left ventricle and into the aorta (cardiac output) and the flow brought into the right atrium and out of the vena cava (cardiac input). These boundary flows necessarily have equivalent mean flow rate and are obtained by fitting the flow profile at the midpoint of the Left Ventricular Outflow Tract and Superior Vena Cava II to that from the Mynard-Smolich model, with the former location used to fit the mean cardiac output/input. Fig. \ref{verfig}A-C shows these derived boundary conditions for the linearized model over one heartbeat as well as the flow from both models at the fitting points. From here, the flow throughout the entire network can be calculated and directly compared to flow data from the Mynard-Smolich model. Fig. \ref{verfig}D and E show the mean and root-mean-square flow for every vessel.

We see strong agreement between the mean flow rates produced by the Mynard-Smolich model and our linearized model throughout the vasculature. Similarly, our linearized model also reproduces the root-mean-squared (RMS) flow well, though with a trend of being slightly larger than that of the Mynard-Smolich model in many vessels. The most notable exceptions to these agreements are the coronary vessels, which undergo significantly more complex dynamics in the Mynard-Smolich via a coupling to the motions of the heart itself. By contrast, our model uses the same dynamic equations for all vessels, an approximation given credence by the fact that the coronary vessels constitute only a small portion of the network. We take the culmination of these results as evidence that our linearized model can appropriately represent the response time and dissipation of the system.

%These comparisons show mostly good agreement between the two models in these metrics with the possible exception of the coronary vessels, which constitute a small portion of the network and undergo significantly more complex dynamics than other vessels in the Mynard-Smolich model. We take the agreement between the mean flows of the Mynard-Smolich model and our linear model as evidence that our linearized model can appropriately represent the response time of the system due to the similar long time steady state. Similarly, the agreement of the root-mean-square flows indicates that our model can accurately account for the dissipation due to similar variances.
%We take the agreement between the mean and root-mean-square flows as evidence that our linearized model can appropriately represent the response time of the system due to the similar long time steady state and dissipation due to similar variances respectively.

Given these verifications, we can proceed to numerically calculate the dissipation and response scaling functions of our linearized network in a method similar to how they were determined in the artificial networks of the previous section. Of note however is that the explicit values of $\bar{\sigma}$ and $\bar{\nu}$ are not calculable for the Mynard-Smolich network due to the capillary beds not being treated as full vessels. As such, the dissipation and response scaling functions cannot be framed as functions of $\bar{\sigma}$ as they were previously, but given that $\bar{\sigma}$ scales as $\sqrt{c_{f}}$ when the compliance of all vessels and capillary bed capacitors are scaled by $c_{f}$, we can use $\sqrt{c_{f}}$ as a proxy measure for $\bar{\sigma}$. Additionally, since $\bar{\nu}$ is independent of $c_{f}$, we can simply use the full response time, $\bar{\nu}\beta$, in place of the nondimensional response scaling function, $\beta$, without altering the form of the dependence on $\sqrt{c_{f}}$.

\begin{figure*}[t]
    \centering
    \includegraphics[width=\textwidth]{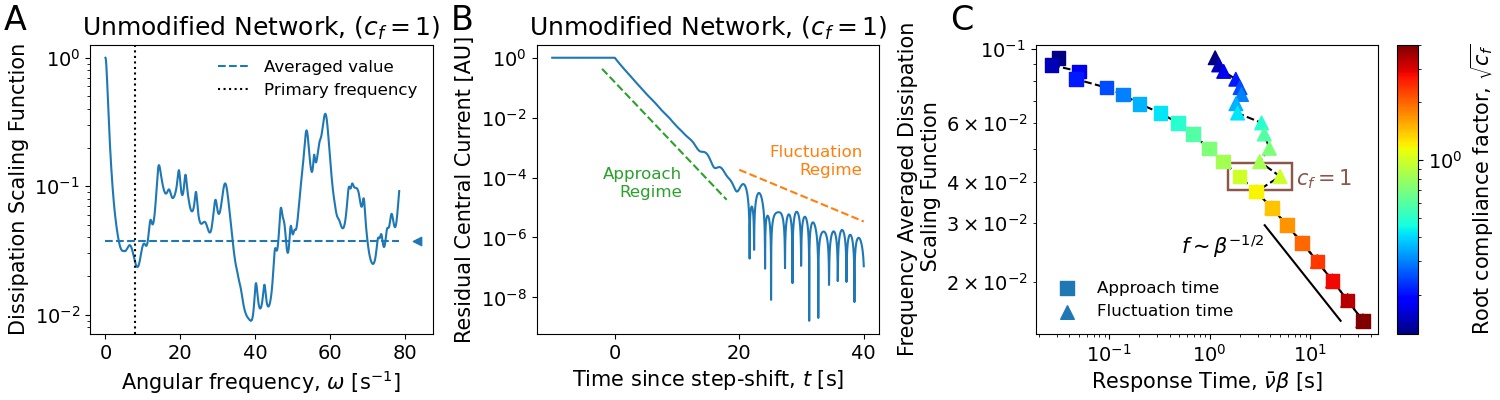}
    \caption{Dissipation and response in the linearized Mynard-Smolich network. A) The dissipation scaling function for the unmodified network up to an angular frequency of 25$\pi$ s$^{-1}$. The dashed blue line and triangle denote the frequency averaged value determined by numeric integration up to $\omega=8000\pi$ s$^{-1}$. Vertical dotted line shows the primary coronary frequency of 75 bpm. B) Response of the unmodified network to a unit magnitude step function boundary condition. The total current through the capillaries exponentially approaches its long time steady state unit value. The residual current is plotted and clearly shows this exponential trend as well as the distinct scalings of the approach and fluctuation regimes. C) Plotting the frequency averaged dissipation scaling function against the approach and fluctuation times as a function of $\sqrt{c_{f}}$ clearly shows the inverse square root power law relation for sufficiently high compliances. The approach and fluctuation times are also seen to merge for values of $c_{f}$ greater than 1.}
    \label{resultsfig}
\end{figure*}

With these reframings, the dissipation scaling function is calculated and frequency averaged in the exact same way as was done in the previous section, producing Fig. \ref{resultsfig}A, but the increased complexity of the network considered in this section required that the response of the system be calculated via a convolution of Fourier transforms rather than a simple simulation (see Supplement). The central current of the network was taken to be the cumulative current through all capillary beds as these are considered to represent the midpoints of the network in same way as the distal nodes did for those depicted in Fig. \ref{diagrams}. Intriguingly, the response of the network to a nonpulsatile shift in boundary current ($H(t)=\Theta(t)$) as monitored from this position displayed a distinct qualitative feature not observed in the simpler artificial networks of the previous section. Fig. \ref{resultsfig}B shows the residual current (the total current subtracted from its long time steady state value of 1) on a semilog scale, from which we see two distinct time scales, rather than the single response time scale seen in the artificial networks. The first regime is a smooth, exponential decay over a time scale denoted as the ``approach time", $\beta_a^{-1}$. The second regime is a series of fluctuations around the long time steady state with magnitudes that decay over a time scale denoted as the ``fluctuation time", $\beta_f^{-1}$. 

This separation of the response into the approach and fluctuation regimes as we have labelled them is not seen in the single vessel case nor the smaller artificial networks represented in Fig. \ref{DVR}, thus making it a distinct aspect of our linearized version of the Mynard-Smolich model. The origin of this phenomenon is unclear based on the results established here and in our previous work \cite{fancher2022mechanical}. The distinct dynamics of the capillary beds, wherein the central current was measured, are unlikely to introduce any new complexities as they are equivalent to a first order approximation of our model in the fluid inertia (see Supplement). We leave further investigations of this to future works.

Regardless of the cause behind these separate response regimes, we still observe the same qualitative behavior relative to dissipation seen in the previously considered artificial networks. Fig. \ref{resultsfig}C plots the dissipation scaling function against both the approach and fluctuation time as a function of $\sqrt{c_{f}}$. The fluctuation time is clearly shown to uphold both the decreasing dissipation and roughly constant fluctuation time for small $\sqrt{c_{f}}$ as well as the inverse square root power law relation at large $\sqrt{c_{f}}$. However, the approach time maintains a consistent scaling with $\sqrt{c_{f}}$ for all values investigated here. Additionally, the two plots are seen to merge and produce equivalent time scales for values of $\sqrt{c_{f}}$ only slightly larger than 1, reinforcing the ubiquity of the tradeoff between energy efficiency and mechanical response time in generic animal vascular networks.

\section{Discussion}
\label{sec:discuss}

In summary, we have shown that compliance in a material transport vessel affects not only the rate of energy dissipation but also the time necessary for the system to respond to changes in boundary conditions. For a single vessel obeying Eqs. \ref{Qdyn} and \ref{Pdyn} with symmetric flow boundary conditions, an explicit tradeoff is seen to exist between the frequency averaged dissipation scaling function, calculated via Eq. \ref{DSFmean}, and the response scaling function. Dissipation scales as the inverse square root of the response time for sufficiently long vessels while short vessels maintain a constant response time with dissipation weakly decreasing with vessel size. This same qualitative relation between dissipation and response was seen to carry into networks of such vessels as well with an identical scaling relation occurring at sufficiently large values of $\bar{\sigma}$. Finally, in constructing a linearized version of the Mynard-Smolich model utilizing the dynamic equations presented here, we observed two distinct response time scales, which we denoted as the approach and fluctuation times. For compliances larger than the unmodified values ($c_{f}=1$), both the approach and fluctuation times obeyed the same inverse square root power law relation seen in the single vessel case and other networks. These times separated for smaller compliances, resulting in the approach time continuously decreasing while the fluctuation time behaved qualitatively similarly to the other considered cases by maintaining a roughly constant value.

The model presented here is of course not the only effort designed to study fluid flow through a network of compliant vessels. The model developed in \cite{flores2016novel} is similarly linear in the current and pressure but does not require the Womersley number to be small nor the fluid to be Newtonian. This allows for a more generalized treatment of the fluid dynamics throughout the network at the cost of having single, well defined space and time scales such as $\lambda$ and $\tau$ and the ability to analytically calculate properties such as the response scaling function. Nonlinear models, such as the Mynard-Smolich model, have also been developed and shown to be better able to reproduce experimentally measured waveforms in the human vascular network specifically \cite{bui2009dynamics,alastruey2011pulse,mynard2015one}. Our linearized model eschews the extreme computational cost and precise experimental fitting traditionally incorporated within these more complex models in favor of a reduced approximation capable of a wide range of analytic solutions in both the frequency and time domains \cite{fancher2022mechanical}. Through an exploration of these solutions, we have observed the existence of a power law tradeoff between dissipation and response time in fluid flow through compliant vessels. This phenomenon is also seen in the network simulations performed here, implying the possibility that it is a robust feature that should manifest in any compliant network of sufficiently large size relative to the inherent wave attenuation length scales. 

%Our linearized adaptation of the Mynard-Smolich network indicates that the human vasculature naturally exists very near the critical point in which the approach and fluctuation time scales merge to produce the inverse square root relation between dissipation and response time, suggesting the further possibility that the compliance of human blood vessels may have evolved to maintain a minimal response time while damping pulsatility as much as possible. Further {\it in silico} experimentation with more detailed, nonlinear models will be necessary to properly support this observation. Additional studies of other animal species might also reveal vasculatures that similarly exist near this critical transition point and elevate this finding to a universal design principle for naturally occurring transport networks.

This work adds to a large body of work on optimization and design principles of biological flow networks, such as \cite{Chang2019,Planck2019,Tekin2016,Postnov2016,Erlich2019}, and much more. In particular, it proposes that the speed with which mechanical information is communicated in the network is an important cost function that needs to be considered alongside other costs such as power dissipation, robustness, perfusion efficiency and adaptability, in determining the optimal vascular topology of large, compliant biological flow systems. The response timescale quantifies the mechanical information propagation cost function; when it is short, it allows the body to respond quickly to changes in the heart rate or other abrupt hemodynamic events and ensures the ability to quickly regulate pressure. We have shown that compliance in a material transport vessel affects both this response timescale as well as the rate of energy dissipation. Minimizing response time by decreasing compliance would result in an increase of the power needed to circulate the blood, establishing a tradeoff. On the other hand, minimizing dissipation by increasing compliance would lead to the attenuation of pressure pulses and elimination of pulsatility altogether. Pulsatility is still present even at the capillary level, although maintaining it comes at an energetic cost, providing support to the argument that the response timescale is an important cost function for the optimal design of animal vascular networks. Understanding this tradeoff is important for understanding the design of human vasculature, but also the comparative physiology and evolution of all vascular systems comprised of elastic walls.

\acknowledgements This research was supported by the NSF Award PHY-1554887 and the Simons Foundation through Award 568888.

\bibliography{refs}

\begin{thebibliography}{10}

\bibitem{lucas2013plant}
W.~J. Lucas, A.~Groover, R.~Lichtenberger, K.~Furuta, S.-R. Yadav,
  Y.~Helariutta, X.-Q. He, H.~Fukuda, J.~Kang, S.~M. Brady, {\em et~al.}, ``The
  plant vascular system: evolution, development and functions f,'' {\em Journal
  of integrative plant biology}, vol.~55, no.~4, pp.~294--388, 2013.

\bibitem{monahan2013evolutionary}
R.~Monahan-Earley, A.~M. Dvorak, and W.~C. Aird, ``Evolutionary origins of the
  blood vascular system and endothelium,'' {\em Journal of thrombosis and
  haemostasis}, vol.~11, pp.~46--66, 2013.

\bibitem{gallego2001optimal}
R.~A. Gallego, A.~J. Monticelli, and R.~Romero, ``Optimal capacitor placement
  in radial distribution networks,'' {\em IEEE Transactions on Power Systems},
  vol.~16, no.~4, pp.~630--637, 2001.

\bibitem{jimenez2010determination}
G.~A. Jim{\'e}nez-Est{\'e}vez, L.~S. Vargas, and V.~Marianov, ``Determination
  of feeder areas for the design of large distribution networks,'' {\em IEEE
  Transactions on Power Delivery}, vol.~25, no.~3, pp.~1912--1922, 2010.

\bibitem{ziari2012optimal}
I.~Ziari, G.~Ledwich, A.~Ghosh, and G.~Platt, ``Optimal distribution network
  reinforcement considering load growth, line loss, and reliability,'' {\em
  IEEE Transactions on Power Systems}, vol.~28, no.~2, pp.~587--597, 2012.

\bibitem{murray1926physiological}
C.~D. Murray, ``The physiological principle of minimum work: I. the vascular
  system and the cost of blood volume,'' {\em Proceedings of the National
  Academy of Sciences of the United States of America}, vol.~12, no.~3, p.~207,
  1926.

\bibitem{sherman1981connecting}
T.~F. Sherman, ``On connecting large vessels to small. the meaning of murray's
  law.,'' {\em The Journal of general physiology}, vol.~78, no.~4,
  pp.~431--453, 1981.

\bibitem{Katifori2010a}
E.~Katifori, G.~J. Sz{\"{o}}llosi, and M.~O. Magnasco, ``{Damage and
  fluctuations induce loops in optimal transport networks.},'' {\em Physical
  Review Letters}, vol.~104, no.~4, p.~048704, 2010.

\bibitem{Postnov2016}
D.~D. Postnov, D.~J. Marsh, D.~E. Postnov, T.~H. Braunstein, N.~H.
  Holstein-Rathlou, E.~A. Martens, and O.~Sosnovtseva, ``{Modeling of Kidney
  Hemodynamics: Probability-Based Topology of an Arterial Network},'' {\em PLoS
  Computational Biology}, vol.~12, no.~7, pp.~1--28, 2016.

\bibitem{Ronellenfitsch2016}
H.~Ronellenfitsch and E.~Katifori, ``{Global Optimization , Local Adaptation ,
  and the Role of Growth in Distribution Networks},'' {\em Phys Rev Lett},
  vol.~117, p.~138301, 2016.

\bibitem{Tekin2016}
E.~Tekin, D.~Hunt, M.~G. Newberry, and V.~M. Savage, ``{Do Vascular Networks
  Branch Optimally or Randomly across Spatial Scales?},'' {\em PLoS
  Computational Biology}, vol.~12, no.~11, pp.~1--28, 2016.

\bibitem{Ronellenfitsch2019}
H.~Ronellenfitsch and E.~Katifori, ``{Phenotypes of Vascular Flow Networks},''
  {\em Physical Review Letters}, vol.~123, no.~24, p.~248101, 2019.

\bibitem{Kirkegaard2020}
J.~B. Kirkegaard and K.~Sneppen, ``{Optimal Transport Flows for Distributed
  Production Networks},'' {\em Physical Review Letters}, vol.~124, no.~20,
  p.~208101, 2020.

\bibitem{Chang2019}
S.~S. Chang and M.~Roper, ``{Microvscular networks with uniform flow},'' {\em
  Journal of Theoretical Biology}, vol.~462, pp.~48--64, 2019.

\bibitem{Planck2019}
F.~J. Meigel, P.~Cha, M.~P. Brenner, and K.~Alim, ``{Robust increase in supply
  by vessel dilation in globally coupled microvasculature},'' {\em Physical
  Review Letters}, vol.~123, no.~22, p.~228103, 2019.

\bibitem{Erlich2019}
A.~Erlich, P.~Pearce, R.~P. Mayo, O.~E. Jensen, and I.~L. Chernyavsky,
  ``{Physical and geometric determinants of transport in fetoplacental
  microvascular networks},'' {\em Science Advances}, vol.~5, no.~4, 2019.

\bibitem{jones1971effects}
E.~Jones, M.~Anliker, and I.~D. Chang, ``Effects of viscosity and constraints
  on the dispersion and dissipation of waves in large blood vessels: Ii.
  comparison of analysis with experiments,'' {\em Biophysical journal},
  vol.~11, no.~12, pp.~1121--1134, 1971.

\bibitem{hashimoto2014central}
J.~Hashimoto, ``Central hemodynamics and target organ damage in hypertension,''
  {\em The Tohoku Journal of Experimental Medicine}, vol.~233, no.~1, pp.~1--8,
  2014.

\bibitem{holenstein1988reverse}
R.~Holenstein and D.~N. Ku, ``Reverse flow in the major infrarenal vessels--a
  capacitive phenomenon,'' {\em Biorheology}, vol.~25, no.~6, pp.~835--842,
  1988.

\bibitem{sherwin2003one}
S.~Sherwin, V.~Franke, J.~Peir{\'o}, and K.~Parker, ``One-dimensional modelling
  of a vascular network in space-time variables,'' {\em Journal of engineering
  mathematics}, vol.~47, no.~3-4, pp.~217--250, 2003.

\bibitem{alastruey2012physical}
J.~Alastruey, T.~Passerini, L.~Formaggia, and J.~Peir{\'o}, ``Physical
  determining factors of the arterial pulse waveform: theoretical analysis and
  calculation using the 1-d formulation,'' {\em Journal of Engineering
  Mathematics}, vol.~77, no.~1, pp.~19--37, 2012.

\bibitem{flores2016novel}
J.~Flores, J.~Alastruey, and E.~C. Poir{\'e}, ``A novel analytical approach to
  pulsatile blood flow in the arterial network,'' {\em Annals of biomedical
  engineering}, vol.~44, no.~10, pp.~3047--3068, 2016.

\bibitem{yigit2016non}
B.~Yigit and K.~Pekkan, ``Non-dimensional physics of pulsatile cardiovascular
  networks and energy efficiency,'' {\em Journal of The Royal Society
  Interface}, vol.~13, no.~114, p.~20151019, 2016.

\bibitem{harazny2014first}
J.~M. Harazny, C.~Ott, U.~Raff, J.~Welzenbach, N.~Kwella, G.~Michelson, and
  R.~E. Schmieder, ``First experience in analysing pulsatile retinal capillary
  flow and arteriolar structural parameters measured noninvasively in
  hypertensive patients,'' {\em Journal of hypertension}, vol.~32, no.~11,
  pp.~2246--2252, 2014.

\bibitem{bui2009dynamics}
A.~Bui, I.~D. {\v{S}}utalo, R.~Manasseh, and K.~Liffman, ``Dynamics of
  pulsatile flow in fractal models of vascular branching networks,'' {\em
  Medical \& biological engineering \& computing}, vol.~47, no.~7,
  pp.~763--772, 2009.

\bibitem{pan2014one}
Q.~Pan, R.~Wang, B.~Reglin, G.~Cai, J.~Yan, A.~R. Pries, and G.~Ning, ``A
  one-dimensional mathematical model for studying the pulsatile flow in
  microvascular networks,'' {\em Journal of biomechanical engineering},
  vol.~136, no.~1, 2014.

\bibitem{perdikaris2015effective}
P.~Perdikaris, L.~Grinberg, and G.~E. Karniadakis, ``An effective fractal-tree
  closure model for simulating blood flow in large arterial networks,'' {\em
  Annals of biomedical engineering}, vol.~43, no.~6, pp.~1432--1442, 2015.

\bibitem{bauerle2020living}
F.~K. B{\"a}uerle, S.~Karpitschka, and K.~Alim, ``Living system adapts
  harmonics of peristaltic wave for cost-efficient optimization of pumping
  performance,'' {\em Physical review letters}, vol.~124, no.~9, p.~098102,
  2020.

\bibitem{fancher2022mechanical}
S.~Fancher and E.~Katifori, ``Mechanical response in elastic fluid flow
  networks,'' {\em Physical Review Fluids}, vol.~7, no.~1, p.~013101, 2022.

\bibitem{mynard2015one}
J.~P. Mynard and J.~J. Smolich, ``One-dimensional haemodynamic modeling and
  wave dynamics in the entire adult circulation,'' {\em Annals of biomedical
  engineering}, vol.~43, no.~6, pp.~1443--1460, 2015.

\bibitem{alastruey2011pulse}
J.~Alastruey, A.~W. Khir, K.~S. Matthys, P.~Segers, S.~J. Sherwin, P.~R.
  Verdonck, K.~H. Parker, and J.~Peir{\'o}, ``Pulse wave propagation in a model
  human arterial network: assessment of 1-d visco-elastic simulations against
  in vitro measurements,'' {\em Journal of biomechanics}, vol.~44, no.~12,
  pp.~2250--2258, 2011.

\end{thebibliography}
\bibliographystyle{ieeetr}

\end{document}